\documentclass{aa}
\usepackage{natbib,twoopt}
\usepackage[breaklinks=true]{hyperref} 
\bibpunct{(}{)}{;}{a}{}{,} 
\makeatletter
\newcommandtwoopt{\citeads}[3][][]{\href{http://adsabs.harvard.edu/abs/#3}%
{\def\hyper@linkstart##1##2{}%
\let\hyper@linkend\@empty\citealp[#1][#2]{#3}}}
\newcommandtwoopt{\citepads}[3][][]{\href{http://adsabs.harvard.edu/abs/#3}%
{\def\hyper@linkstart##1##2{}%
\let\hyper@linkend\@empty\citep[#1][#2]{#3}}}
\newcommandtwoopt{\citetads}[3][][]{\href{http://adsabs.harvard.edu/abs/#3}%
{\def\hyper@linkstart##1##2{}%
\let\hyper@linkend\@empty\citet[#1][#2]{#3}}}
\newcommandtwoopt{\citeyearads}[3][][]%
{\href{http://adsabs.harvard.edu/abs/#3}
{\def\hyper@linkstart##1##2{}%
\let\hyper@linkend\@empty\citeyear[#1][#2]{#3}}}
\makeatother
\usepackage{graphicx}
\usepackage{amsmath}
\usepackage[varg]{txfonts}
\graphicspath{{../pictures/}}

\begin{document}

\title{Mean gas opacity for circumstellar environments and equilibrium temperature degeneracy}

\author{M.~G.~Malygin\inst{1,2}, R.~Kuiper\inst{1,3}, H.~Klahr\inst{1}, C.~P.~Dullemond\inst{4}, Th.~Henning\inst{1}}

   \institute{Max-Planck-Institut f\"ur Astronomie, K\"onigstuhl 17, D-69117 Heidelberg, Germany\\
              \email{malygin@mpia.de}
         \and
     Fellow of the International Max-Planck Research School for Astronomy and Cosmic Physics 
     at the University of Heidelberg (IMPRS-HD)
         \and
     Universit\"at T\"ubingen, Institut f\"ur Astronomie und Astrophysik, Auf der Morgenstelle 10, D-72076 T\"ubingen, Germany
         \and
     Zentrum f\"ur Astronomie der Universit\"at Heidelberg, Institut f\"ur Theoretische Astrophysik, Albert-Ueberle-Stra{\ss}e 2, 69120 Heidelberg, Germany
             }

   \date{Received date; accepted date}

 
  \abstract
   { 
       In a molecular cloud dust opacity typically dominates over gas opacity, yet in the vicinities of forming stars dust is depleted, and gas is the sole provider of opacity.
       In the optically thin circumstellar environments the radiation temperature cannot be assumed to be equal to the gas temperature, hence the two-temperature Planck means are necessary to calculate the radiative equilibrium. 
   }
   {
       By using the two-temperature mean opacity one does obtain the proper equilibrium gas temperature in a circumstellar environment, which is in a chemical equilibrium. 
       A careful consideration of a radiative transfer problem reveals that the equilibrium temperature solution can be degenerate in an optically thin gaseous environment. 
   }
   {
       We compute mean gas opacities based on the publicly available code DFSYNTHE by Kurucz and Castelli. 
       We performed the calculations assuming local thermodynamic equilibrium and an ideal gas equation of state.
       The values were derived by direct integration of the high-resolution opacity spectrum. 
   }
   {
       We produced two sets of gas opacity tables: Rosseland means and two-temperature Planck means\thanks{The tables will be available via \href{http://cdsweb.u-strasbg.fr/}{http://cdsweb.u-strasbg.fr/} as well as via \href{http://www.mpia-hd.mpg.de/homes/malygin}{http://www.mpia.de/\string~malygin}.}.   
       For three metallicities $\mathrm{[Me/H]} = 0{.}0,\pm0{.}3$ we covered the parameter range $3{.}48\leq \log T_\mathrm{rad}\mathrm{~[K]} \leq 4{.}48$ in radiation temperature, $2{.}8\leq\log T_\mathrm{gas}\mathrm{~[K]}\leq6{.}0$ in gas temperature, and $-10 \leq \log P \mathrm{~[dyn\,cm^{-2}]} \leq 6$ in gas pressure. 
       We show that in the optically thin circumstellar environment for a given stellar radiation field and local gas density there are several equilibrium gas temperatures possible. 
   }
   {
       We conclude that, in general, equilibrium gas temperature cannot be determined without treating the temperature evolution. 
   }


   \titlerunning{Mean gas opacity and equilibrium temperature degeneracy}
   \authorrunning{M.G.~Malygin et al.}
   \maketitle
%


\section{Introduction}\label{sec:INTRO}
Mean gas opacities without a dust continuum contribution have many applications in astrophysics \citep[e.g.][]{HEL00,FREED08,HEL09}. 
It will manifest its dominance in a dust-depleted medium resulting either from a low-metallicity environment or when the equilibrium temperature gets higher than the local dust sublimation temperature. 
Therefore, it plays a crucial role in stellar physics, but is also relevant for processes related to stellar feedback in the circumstellar medium (CSM). 
Gas opacity has to be taken into account when describing inner regions of accretion disks \citep{MUZEROLLE04,VAIDYA09,ZHANG11}, cooling of non-accreting hot stellar remnants \citep[e.g.][]{ROHR12}, dust formation around pulsating AGB stars \citep{SCHIRR03}, or calculating energy balance of Type Ia supernovae \citep[e.g.][]{DESS13}. 
There is evidence that the inner gaseous parts of disks around Herbig stars may even be optically thick \citep{ISELLA06}; the same is expected for accretion disks around forming massive stars \citep{ZHANG11,TN11,KUI13}. 
Thus, although on molecular-cloud scales dust opacity dominates over gas opacity, the various dust-free regions play a vital role in the stellar feedback processes that are essential for shaping and developing the ISM and which are important for linking stellar and cosmological scales \citep[][]{PELUPE09}. 
Last but not least, emission from molecular species in the gas phase are important probes of the gas properties and the kinematics of star-forming regions. 

The high angular-resolution observations with the Atacama Large Millimeter Array are now pushing the field of numerical modelling to resolve the circumstellar gaseous interiors and to properly treat gas opacity there. 

A typical situation in a circumstellar environment is that a colder gas is irradiated by a hotter source: $T_\mathrm{gas} < T_\mathrm{rad}$. 
In this case, the absorption from the gas as expressed in terms of the Planck mean opacity will depend on a) the gas temperature determining its chemical state and the level populations and b) the radiation temperature defining the energy distribution of the ambient radiation field via the Planck function: $\kappa_\mathrm{P} = \kappa_\mathrm{P}\left(T_\mathrm{gas},\rho,T_\mathrm{rad}\right)$. 
We shall from now on refer to this as a two-temperature Planck mean. 
We argue that this mean is essential for the equilibrium temperature determination in an optically thin gas (see Sect.~\ref{sec:OTG}).
If $T_\mathrm{rad} = T_\mathrm{gas}$, it turns into the common single-temperature Planck mean $\kappa_\mathrm{P}\left(T_\mathrm{gas},\rho\right)$.
The Rosseland mean $\kappa_\mathrm{R}\left(T_\mathrm{gas},\rho\right)$ is introduced to provide radiation transport in an optically thick medium, where $T_\mathrm{rad} = T_\mathrm{gas}$. 
We will use the following designations throughout the paper: $\kappa_\mathrm{P}\left(T_\mathrm{gas},\rho,T_\mathrm{rad}\right) = \kappa_\mathrm{P}\left(T_\mathrm{gas},T_\mathrm{rad}\right)$ for the two-temperature Planck mean, $\kappa_\mathrm{P}\left(T_\mathrm{gas},\rho\right) = \kappa_\mathrm{P}\left(T_\mathrm{gas}\right)$ for the Planck mean, and $\kappa_\mathrm{R}\left(T_\mathrm{gas},\rho\right) = \kappa_\mathrm{R}\left(T_\mathrm{gas}\right)$ for the Rosseland mean. 
We will drop $\rho$ for sake of brevity. 

The currently-available gas opacity tables \citep{BELLLIN94,HEL00,SEM03,FERG05,FREED08,HEL09} do not include two-temperature Planck mean opacities. 
We compute new gas opacity tables with the two-temperature Planck means and Rosseland means. 
In Sect.~\ref{sec:GOP} we give a brief description of the calculations. 
In Sect.~\ref{sec:GOPRES} we present the new gas opacity tables. 
In Sect.~\ref{sec:OTG} we compute radiative equilibria in an optically thin gas and explain the degeneracy of the equilibrium temperature solution.
We discuss results in Sect.~\ref{sec:DISCUS}.
Summary and outline of future applications are given in Sect.~\ref{sec:SUM}.


\section{Gas opacity calculations}\label{sec:GOP}

\subsection{Motivation}\label{sec:GOP:MOTIV}
The are two motivations for calculating a new gas opacity. 
First, the gas opacity tables available \citep{SEATON94,BELLLIN94,HEL00,SEM03,FERG05,FREED08,HEL09} do not include two-temperature Planck means, which we argue to be essential for our application cases. 
Second, \citet{FERG05} compared single-temperature Planck means obtained by different groups and found a disagreement at gas temperatures $\log T_\mathrm{gas} \ga 3{.}12$: only data from \citet{FERG05} and \citet{SEATON94} agree at atomic temperatures $\log T_\mathrm{gas}\ga 3{.}48$ \citep[see Fig.~12 in][]{FERG05}. 
The discrepancy at lower gas temperatures originates from insufficient frequency sampling in earlier studies, although some contributions might have come from utilising different molecular line lists and/or equations of state. 
The two factors suggest that an independent computation of gas opacity is justified. 

\subsection{Procedure}\label{sec:GOP:PROC}
We adopt the open-source code DFSYNTHE\footnote{Available at \href{http://wwwuser.oat.ts.astro.it/castelli/sources/dfsynthe.html}{http://wwwuser.oat.ts.astro.it/castelli/sources/dfsynthe.html}} by Kurucz and Castelli \citep[described in][]{CAST05}. 
Originally, the code was intended to calculate opacity probability distribution functions \citep[][]{KUR70, KUR74}, which means the opacity spectrum $\kappa_\nu$ is determined during the computation.
Here we implement the calculation of Rosseland and two-temperature Planck means.

We tabulate the means for a broad range of gas densities ($10^{-20} - 10^{-2}\mbox{ g\,cm}^{-3}$), gas temperatures ($700-10^{6}\mbox{ K}$), and for three different metallicities ($-0{.}3$, $0{.}0$, $0{.}3$~dex). 
For the two-temperature Planck means we considered $3\,000\mbox{ K}<T_\mathrm{rad}<30\,000\mbox{ K}$. 
If needed, a wider parameter range can be considered. 

All the calculations were done assuming equilibrium chemistry governed by local thermodynamic equilibrium (LTE), and an ideal gas equation of state. 
The reference values for solar metallicity atomic abundances were taken from~\citet{GREV98}. 

There is a list of input parameters the user should supply: atomic abundances, abundance scaling factors (metallicities), gas pressures, gas temperatures, and radiation temperatures. 
The code at first calculates equilibrium abundances of compounds and equilibrium level populations. 
Using those, it works through the line lists to calculate the line absorption, which is added to the continuum absorption to get the total opacity spectrum. 
Finally, the requested forms of frequency averages are calculated for each set of input parameters.  

The line and continuum opacity data are taken from Kurucz's CD-ROMs Ns 1, 15, and 20-24\footnote{\href{http://kurucz.harvard.edu/cdroms.html}{http://kurucz.harvard.edu/cdroms.html}} \citep{KUR93a,KUR93b}. 
Among the compounds included are $\mbox{H}_2\mbox{O}$, TiO, CO, CO$_2$, H$_2$S, $\mbox{H}_2$, NH, HF, MgH, AlH, SiH, HS, HCl, C$_2$, CN, AlC, SiC, CS, N$_2$, NO, MgN, AlN, NH$_2$, NS, O$_2$, MgO, AlO, SiO, SO, CaO, FeO, MgS, AlS, SiS, S$_2$, CaS, FeS, C$_2$H, CNH, COH, NOH, OOH, CH$_2$. 
We compile the water line list based on BT2 data \citep{BT2}; the TiO line list was taken from \citet{SCHWENKE98} (CD-ROM No 24). 
All the line lists were re-sampled at a resolving power of $R=\lambda/\delta\lambda = 500\,000$. 
At this resolution ($\ga3{.}5\times10^{6}$ wavelength points) the opacity spectrum is calculated over the wavelength range of $8{.}9\mbox{ nm}$--$10\mbox{ }\mu\mbox{m}$. 
The Rosseland mean can be equally well calculated with even a factor of $\sim100$ lower resolution, because it is dominated by line wings (broad regions), but the Planck mean is dominated by line cores (narrow regions), which demands the Voigt profiles to be properly resolved and sampled (see Sect.~\ref{sec:RES:COMPAR}). 

For all species of the line data, the line absorption coefficient is calculated on a given frequency grid according to the formula \citep[see][]{CAST05} 
\begin{equation}\label{eq:lnu}
    \begin{split} 
    l_{\nu} = \frac{1}{\rho}\frac{\sqrt\pi \mathrm{e}^2}{m_\mathrm{e} c}f_{ij}
    \frac{g_iN_j}{U_j}\exp\left(\chi_i/kT\right)\times\frac{1}{\Delta\nu_\mathrm{D}}\times\\ 
    H\left(\frac{\Gamma}{4\pi\Delta \nu_\mathrm{D}},\frac{\Delta \nu}{\Delta \nu_\mathrm{D}}\right)\times\left(1-\exp\left(-h\nu/kT\right)\right), 
    \end{split} 
\end{equation}
where $\rho$ is the gas mass density, 
    $\mathrm{e}$   is the charge of an electron, 
    $m_\mathrm{e}$   is the electron mass, 
    $c$      is the speed of light; 
    $f_{ij}$ corresponds to the Ladenburg $f$ or oscillator strength, 
    $g_i$    is statistical weight of the lower level $i$,
    $N_j$    is number density of the absorbing species in the $j$ ionisation stage,
    $U_j$    is partition function of the absorbing species in the $j$ ionisation stage, 
    $\chi_i$ is ionisation potential from the lower $i$ level; 
    $\Delta\nu_\mathrm{D} = \nu_0/c\sqrt{2kT/\mu+\xi_\mathrm{t}^2}$ is the total Doppler broadening of the transition at the $\nu_0$ frequency, with 
    $k$ being the Boltzmann constant and  
    $\mu$ the mean molecular weight, 
    $\xi_\mathrm{t}$ the additional non-thermal broadening parameter, e.g. due to turbulent motions, usually called microturbulence parameter. 
    The Voigt function is given by
    \begin{equation}
    H\left(a,v\right) = \dfrac{a}{\pi} \int_{-\infty}^{+\infty} \dfrac{\exp\left(-y^{2}\right)dy}{\left(v-y\right)^{2}+a^{2}} 
    \label{eq:voigt}
    \end{equation}
    with 
    $a = \Gamma/(4\pi\Delta \nu_\mathrm{D})$ and
    $v = \Delta \nu / \Delta \nu_\mathrm{D}$;  
    $\Gamma = \Gamma_\mathrm{L} + \Gamma_\mathrm{S} + \Gamma_\mathrm{W}$ is a sum of Lorentz, Stark, and van der Waals broadening. 
The code takes collisional broadening by atomic and molecular hydrogen and neutral helium into account \citep{KUR81}. 
We set $\xi_\mathrm{t} = 0$, facilitating comparison to other data (see Sect.~\ref{sec:RES:COMPAR}). 

Total absorption is a sum of line and continuum absorptions from all the species: $\kappa_\nu = l_{\nu} + \sigma_{\nu}$. 
From the frequency-dependent opacities $\kappa_\nu$ we calculate the Rosseland mean
\begin{equation} \label{eq:ross}
    \kappa^{-1}_\mathrm{R}\left(T_\mathrm{gas}\right) = \dfrac{\int\kappa_\nu^{-1}\left(T_\mathrm{gas}\right)\times\partial_{T}B_\nu\left(T_\mathrm{gas}\right) d\nu}{\int\partial_{T}B_\nu\left(T_\mathrm{gas}\right) d\nu},
\end{equation} 
the two-temperature Planck mean 
\begin{equation}\label{eq:ttpl}
    \kappa_\mathrm{P}\left(T_\mathrm{gas}, T_\mathrm{rad}\right) = \dfrac{\int{\!\kappa_\nu(T_\mathrm{gas})\times B_{\nu}(T_\mathrm{rad})d\nu}}{\int B_{\nu}(T_\mathrm{rad})d\nu}, 
\end{equation} 
and the single-temperature Planck mean 
\begin{equation}\label{eq:pl}
    \kappa_\mathrm{P}\left(T_\mathrm{gas}\right) = \kappa_\mathrm{P}\left(T_\mathrm{gas},T_\mathrm{rad} = T_\mathrm{gas}\right) = \dfrac{\int{\!\kappa_\nu(T_\mathrm{gas})\times B_{\nu}(T_\mathrm{gas})d\nu}}{\int B_{\nu}(T_\mathrm{gas})d\nu}. 
\end{equation} 
In the following section we present the results of the calculations.


\section{Results}\label{sec:GOPRES} 
In this section all the values of gas opacity are given for solar atmospheric atomic abundances \citep{GREV98} and zero turbulent broadening unless otherwise stated. 
However, different authors used different references for solar abundances. 
We will focus on the single-temperature and the two-temperature Planck mean because these are the points of major improvement in comparison to previous studies. 

\subsection{Single-temperature Planck mean}\label{sec:RES:PLSURF}
\begin{figure}[!htb]
  \begin{center}
   \includegraphics[width=1.0\linewidth, angle=0]{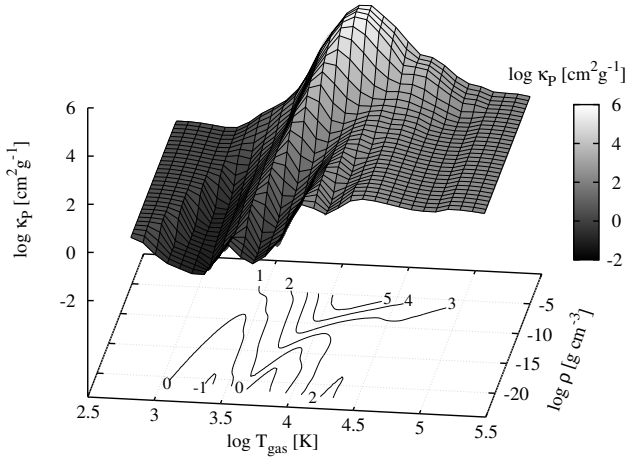}
\caption{
    Surface of the Planck mean gas opacity in logarithmic parameter space. 
    The contours on the xy-plane are iso-contours $\log\kappa_\mathrm{P} = \mathrm{const}$, the labels giving the values.  
    The colour scale is the same as the height-scale. 
    The grid is superimposed only for better rendering and does not reproduce the actual computational grid, which is irregular.  
    See text for more details.
\label{fig:PLANCKSURF}}  
\end{center} 
\end{figure}

Surface plot in Fig.~\ref{fig:PLANCKSURF} shows the single-temperature Planck mean gas opacity as a function of gas temperature and density. 
The iso-contours reveal regions with steep and shallow gradients of $\kappa_\mathrm{P}$ with respect to these parameters. 
These are caused by changes in equilibrium abundance of the species, level populations in atoms and molecules and by the shift of the maximum of the weighting Planck function. 
We briefly highlight the processes that shape the surface. 
At gas temperatures $\la 1\,000\mbox{ K}$, a ubiquity of numerous molecular lines from different species make different parts of an opacity spectrum within the correspondent wavelength region of $1-50\mbox{ }\mu\mbox{m}$ look alike, so that the frequency-averaged values vary moderately with temperature and density. 
This is responsible for the flattish region in the top-left corner of the surface. 
In the temperature range from $1\,000\mbox{ K}$ to $3\,000\mbox{ K}$ there is a gap in opacity because none of the species that comprise the gas mixture has prominent bands in the corresponding wavelength region. 
The gap is pronounced at low densities, but is filled at the highest densities because of enormous pressure-broadening of the lines, in particular those from alkali atoms \citep[see Sect.~\ref{sec:RES:CON} in the current paper, as well as][]{FREED08}. 

The rising opacity from H$^-$, H, and metals results in the high bump, which peaks at $\sim10^{4}\mbox{ K}$. 
The mean opacity on the top of the ``mountain'' is governed by H bound-bound (b-b), bound-free, and free-free absorption. 
The contribution from atomic metals at these temperatures comes at higher frequencies and therefore is exponentially suppressed by the Wien cut.
At higher temperatures the ionisation of hydrogen eliminates its b-b source of opacity and the mean opacity starts to decline. 
To the right of the larger hydrogen hump there is a smaller helium hump. 
Free-free and metal line absorption dominate the mean at higher temperatures up to $10^{6}\mbox{ K}$. 
At these temperatures the Planck mean drops moderately with density, but is sensitive to metallicity.
Finally, it settles at the wavelength-independent Thomson scattering cross-section, which results in a plateau at temperatures $\ga 10^{6}\mbox{ K}$ (not shown in the plot). 

\subsection{Two-temperature Planck mean}\label{sec:RES:TTPL}
\begin{figure}[!htb]
  \begin{center} 
     \includegraphics[width=1.0\linewidth, angle=0]{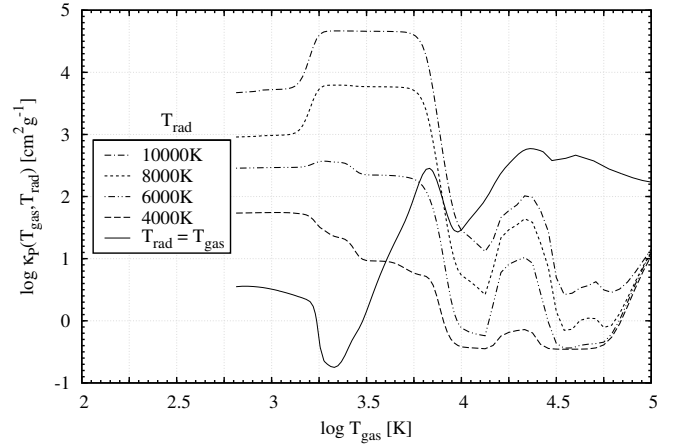} 
    \caption{ Two-temperature Planck mean opacity as a function of gas temperature. 
	The plot is for a constant density of $10^{-13}\mbox{ g\,cm}^{-3}$ and for different radiation temperatures $T_\mathrm{rad}$, as indicated in the legend. 
    The single-temperature Planck mean $\left(T_\mathrm{rad}=T_\mathrm{gas}\right)$ at the same density is given in solid.
} 
    \label{fig:TTPL} 
    \end{center}
\end{figure}
Figure~\ref{fig:TTPL} shows the two-temperature Planck mean as a function of gas temperature for several radiation temperatures and at a constant density of $10^{-13}\mbox{ g\,cm}^{-3}$. 
It is clearly seen from the plot that $\kappa_\mathrm{P}(T_\mathrm{gas}, T_\mathrm{rad} > T_\mathrm{gas}) > \kappa_\mathrm{P}(T_\mathrm{gas} = T_\mathrm{rad})$. 
This is because the parts of the opacity spectrum, which get exponentially cut by the Wien factor in the single-temperature Planck mean ($T_\mathrm{gas} = T_\mathrm{rad}$), are becoming pronounced by the Planck function with higher temperature in the two-temperature Planck mean.  
Consequently, at a constant gas temperature, $T_\mathrm{gas}$, $\kappa_\mathrm{P}\left(T_\mathrm{gas}, T_\mathrm{rad}\right)$ is higher for higher radiation temperatures. 
The ratio $\kappa_\mathrm{P}\left(T_\mathrm{gas}\right)/\kappa_\mathrm{P}\left(T_\mathrm{gas}, T_\mathrm{rad}\right)$ differs from $\kappa_\mathrm{P}\left(T_\mathrm{gas}\right)/\kappa_\mathrm{P}\left(T_\mathrm{rad}\right)$, especially at $T_\mathrm{rad}\ga5\,700\mbox{ K}$. 
The qualitative look of the plot remains the same for other densities. 

\subsection{Metallicity}\label{sec:RES:MEH}
\begin{figure}[htb]
  \begin{center}
   \includegraphics[width=1.0\linewidth, angle=0]{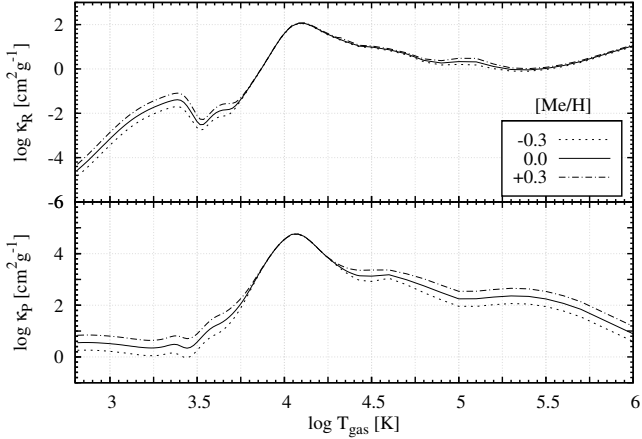}
   \caption{ Rosseland (top panel) and single-temperature Planck (bottom panel) means versus gas temperature. 
    Logarithms of the values are plotted for $\rho = 10^{-8}\mbox{ g\,cm}^{-3}$ at three different metallicities, indicated in the legend. 
\label{fig:KPKRMEH}}  
\end{center} 
\end{figure}

Figure~\ref{fig:KPKRMEH} presents the Rosseland and the single-temperature Planck means at a constant gas density of $10^{-8}\mbox{ g\,cm}^{-3}$ as functions of gas temperature for three different metallicities: [Me/H] = $-0{.}3$, $0{.}0$, and $0{.}3$. 
Other metallicities may be readily added on demand. 

Varying metallicity affects the means at high temperatures via atomic metal absorption and at low temperatures via change in the contents of the compounds. 
In between, the opacity is dominated by hydrogen bound-bound, bound-free and free-free absorption (see Sect.~\ref{sec:RES:PLSURF}) and is insensitive to the amount of metals.  
The hydrogen temperatures $5\,700 - 17\,700\mbox{ K}$  are clearly distinguished in each panel. 
At molecular temperatures $\la 5\,700\mbox{ K}$ both means show the dependence on metallicity, but at metallic temperatures $\ga17\,700\mbox{ K}$ the Planck mean depends more strongly on the metal abundance than the Rosseland mean does.
This is because the Planck mean is being dominated by line cores from metals in this temperature regime. 

\subsection{Connection to low temperature atmospheric opacity}\label{sec:RES:CON}
\citet{FREED08} calculated the gas opacity relevant for high gravity environments, where formation and settling of condensates into clouds occur and considered a mixture of only 11 molecules and 5 alkali elements: H$_{2}$O, TiO, CO, CO${}_{2}$, H${}_{2}$S, CH${}_{4}$, NH${}_{3}$, PH${}_{3}$, VO, FeH, CrH, and Li, Na, K, Cs, Rb, respectively. 
In contrast, our calculations apply for low gravity equilibrium gas chemistry with no settling. 
We include more than 40 compounds, five of which overlap with the first five listed above (see Sect.~\ref{sec:GOP:PROC}), as well as such alkali atoms as Li, Na, and K, but not Cs and Rb.  
We extend the parameter space up to $10^{-2}\mbox{ g\,cm}^{-3}$ to compare the two studies.

\begin{figure}[!htb]
  \begin{center} 
      \includegraphics[width=1.0\linewidth, angle=0]{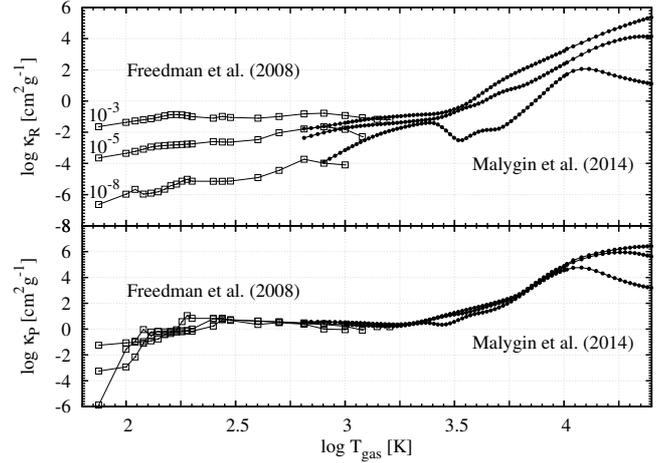} 
      \caption{ Mean gas opacity: comparison with \citet{FREED08}. 
  The Rosseland means (top panel) and the Planck means (bottom panel) as a function of gas temperature. 
      The open squares are data from \citet{FREED08}, the filled circles are from this work. 
      Each curve corresponds to a constant gas density $\rho[\mbox{ g\,cm}^{-3}]=10^{-3}$, $10^{-5}$, $10^{-8}$ as labelled in the upper panel.
      In the lower panel the different lines are difficult to distinguish because the Planck mean is less sensitive to density.
    } \label{fig:FREEDMAN} 
    \end{center}
\end{figure}

Figure~\ref{fig:FREEDMAN} shows a connection between \citet{FREED08} and this work. 
The iso-density contours of the Rosseland mean opacity from the two works cross each other. 
This indicates that there is an overlap region, where both determine similar opacity values, despite the fact that the included species differ significantly. 
The intersection temperatures of the Rosseland means increase with density: $650\mbox{ K}$ (the lowest temperature in our calculations) for $10^{-8}\mbox{ g\,cm}^{-3}$, $1\,400\mbox{ K}$ for $10^{-3}\mbox{ g\,cm}^{-3}$, and $4\,000\mbox{ K}$ (the upper boundary of Freedman's calculations) for $10^{-2}\mbox{ g\,cm}^{-3}$. 

The Planck mean is to a high degree density-independent in the region of overlap: the iso-density contours pile up at about one value ($\log\kappa_\mathrm{P} \approx 0{.}5$).
The intersection temperatures for densities $\la10^{-4}\mbox{ g\,cm}^{-3}$ correspond to the lower boundary of our calculations ($650\mbox{ K}$), and get higher at the highest densities ($1\,500\mbox{ K}$ and $10\,000\mbox{ K}$ at $10^{-3}$ and $10^{-2}\mbox{ g\,cm}^{-3}$, respectively). 

To explain this we have to consider the difference in the species included in the two calculations and establish the temperature ranges of importance for each group of those species. 
The species CH$_4$, NH$_3$, PH$_3$, CrH, FeH, and VO, which are not included in our calculations although they are in \citet{FREED08}, have prominent bands within $\sim5-1\,000\mbox{ }\mu\mbox{m}$ \citep[see Fig.~4 in][and references therein]{FREED08}. 
Accordingly, they contribute to the mean opacity at temperatures $\la 1\,000\mbox{ K}$. 
The species included in each calculation, H$_2$O, TiO, CO, CO$_2$, and H$_2$S, are relevant absorbers at temperatures $\ga 1\,000\mbox{ K}$ because they have prominent bands within $\sim0{.}1-5\mbox{ }\mu\mbox{m}$. 
In the intermediate region ($1-5\mbox{ }\mu\mbox{m}$) the opacity from alkali atoms Li, Na, K, Cs, and Rb is prominent, especially at high pressures. 
\citet{FREED08} included more alkali species (Cs, Rb) and applied a more sophisticated collisional broadening prescription, suited for pressures as high as $10^{7} - 10^{8}\mbox{ dyn\,cm}^{-2}$ \citep{BURR00,BURR03}.
As inferred from those works, the pressure-enhanced far-wings from alkali atoms make a significant contribution to the Rosseland mean at temperatures about $1\,000 - 1\,500\mbox{ K}$. 
That is why the Rosseland means intersect within this region and the value of the intersection temperature increases with density. 
The Planck mean is dominated by line cores, and is less sensitive to pressure: a contribution from the alkali atoms appears only at densities as high as $\ga10^{3}\mbox{ g\,cm}^{-3}$. 

The above reasoning is a qualitative analysis of the connection between the two data sets. 
A quantitative consideration would have involved a comparison between the prescriptions of collisional broadening and accounted for the effects of relative abundances.
We simply remark that the intersection temperatures would be somewhat higher should we decrease in our calculations the abundance of the common species bearing the same atoms as those composing the remaining species in \citet{FREED08}. 

\subsection{Comparison to other studies}\label{sec:RES:COMPAR}
\begin{figure}[!htb] 
  \begin{center} 
    \includegraphics[width=1.0\linewidth, angle=0]{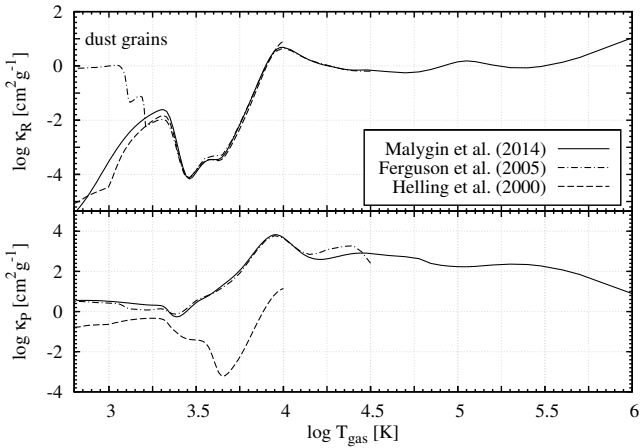} 
    \caption{ Mean opacity: comparison to other studies. 
	The Rosseland (top panel) and the Planck (bottom panel) means at a constant gas density of $10^{-10}\mbox{ g\,cm}^{-3}$. 
	This work -- solid line, \citet{HEL00} -- dashed, and \citet{FERG05} -- dot-dashed. 
	Data from \citet{FERG05} contain opacity from dust grains. 
	At lower temperatures the Rosseland average from the dust opacity dominates the Rosseland average from the gas opacity, but the Planck mean of the dust is comparable to the Planck mean of the gas. 
    } \label{fig:COMPAR} 
    \end{center}
\end{figure}
Figure~\ref{fig:COMPAR} shows a comparison of mean opacities as obtained by different groups. 
The top panel shows the Rosseland mean, and the bottom panel the Planck mean. 
Each panel corresponds to the same density of $10^{-10}\mbox{ g\,cm}^{-3}$. 
Data from \citet{FERG05} include opacity from dust grains. 
Data from both \citet{HEL00} and this work are pure gas opacity. 
Nonetheless, our Planck means compare to dust opacity at temperatures below $1\,500\mbox{ K}$: high absorption in line cores is able to provide the Planck average, comparable to that from continuum opacity of typical ISM dust. 
The Rosseland mean opacity of the dust, however, dominates over Rosseland mean gas opacity at low temperatures. 

We note a good agreement with \citet{FERG05} in both means. 
However, our calculations cover a wider parameter space in terms of temperature and density. 
To the best of our knowledge, this is the largest set of mean gas opacities obtained within one framework, using a single computer code. 

Because of the improper frequency sampling, Planck means from \citet{HEL00} are significantly lower than ours or those from \citet{FERG05}. 
Whereas the number of randomly picked frequencies was sufficient to produce a good approximation for the Rosseland mean values, this number is not sufficient for Planck mean averages, because Rosseland means are harmonic averages while Planck means are direct averages. 
There are not enough random frequency points to sample all absorption features sufficiently. 
This error is smaller for the Rosseland mean, but is substantial for the Planck mean. 
Because of the random\footnote{actually, pseudo-random with respect to positions of the absorption features} frequency sampling, the uncertainties introduced into the Planck opacities by under-sampling are not systematic, but are instead of a stochastic nature. 
\begin{figure}[!htb] 
  \begin{center} 
    \includegraphics[width=1.0\linewidth, angle=0]{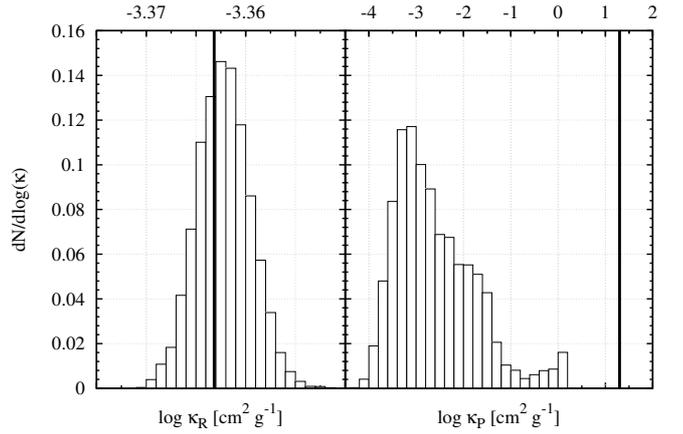}
    \caption{ Normalised distributions of logarithms of Rosseland (left) and Planck (right) means. 
	The values were determined at $4\,466\mbox{ K}$ and $10^{-10}\mbox{ g\,cm}^{-3}$ and calculated from $5\times10^{3}$ sampling points randomly spread over the spectrum ($8{.}9\mbox{ nm} - 10\mbox{ }\mu\mbox{m}$). 
    Total number of trials is $10^{4}$. 
    The fiducial values (those calculated from $\sim3{.}5\times10^{6}$ sampling points) $\log\kappa_\mathrm{R} = -3{.}36317$ and $\log\kappa_\mathrm{P} = 1{.}296$ are marked with thick vertical lines.
    We note the different scales on the $x$-axes.
    }\label{fig:kappadistr} 
    \end{center}
\end{figure}
To verify this statement, we choose the temperature of the biggest disagreement in single-temperature Planck means between \citet{HEL00} and this work at the density of $10^{-10}\mbox{ g\,cm}^{-3}$ ($\approx4\,466\mbox{ K}$; see Fig.~\ref{fig:COMPAR}) and perform $10^{4}$ calculations of Planck and Rosseland means each with $5\times10^{3}$ random frequencies. 
The resulting distributions of $10^{4}$ individually calculated Rosseland and Planck means are shown in Fig.~\ref{fig:kappadistr}. 

The distribution of the Rosseland mean is Gaussian-like, narrow, and peaks at the fiducial value of $\log\kappa_\mathrm{R}=-3{.}36317$.  
Instead, the distribution for the Planck mean is asymmetric, broad (encompassing four orders of magnitude), and lies below the fiducial value of $\log\kappa_\mathrm{P} \approx 1{.}296$. 
This behaviour is exclusively due to the difference in the averaging process: the Rosseland mean is attracted by low-opacity regions between the lines, whereas the Planck mean is attracted by the line cores. 
The regions between the lines are much wider than the line cores, yielding a higher chance to sample from these regions. 
The most probable values of the two distributions are almost equal ($\approx-3{.}3$) because as long as the lines are neglected, the continuum opacity is essentially flat around the Planck function maximum at the considered temperature. 
The Planck mean obtained in \citet{HEL00} for the given temperature and density, $\log\kappa_\mathrm{P}\approx-3{.}2$, is close to this value.


\section{Equilibrium temperature degeneracy in an optically thin gas}\label{sec:OTG}
In this section we calculate radiative equilibrium in an optically thin gas and explain the equilibrium temperature degeneracy. 
\subsection{Illustrative example} 
The condition of radiative equilibrium in an optically thin gas, where local net flux of radiation is $\left|\vec{F_\mathrm{rad}}\right| = \int d\nu\oint d\vec{\Omega} \vec{F}_{\nu}$, reads
\begin{equation}
    \kappa_\mathrm{P}\left(T_\mathrm{gas}\right)\times aT^4_\mathrm{gas} = \kappa_\mathrm{F}\times\frac{\left|\vec{F_\mathrm{rad}}\right|}{c},
    \label{eq:radeq}
\end{equation}
where $a$ is the radiation constant and $\kappa_\mathrm{F} = \left|\vec{F_\mathrm{rad}}\right|^{-1} \int \kappa_{\nu} |\vec{F}_{\nu}| d\nu$ is the flux averaged opacity. 
If the source of radiation is a black body with temperature $T_\mathrm{rad}$ and the gas is in LTE with $T_\mathrm{gas}$, $\kappa_\mathrm{F}$ becomes the two-temperature Planck mean: 
\begin{equation}
    \kappa_\mathrm{P}\left(T_\mathrm{gas}\right)\times aT_\mathrm{gas}^4 = \kappa_\mathrm{P}\left(T_\mathrm{gas}, T_\mathrm{rad}\right) \times \frac{\left|\vec{F_\mathrm{rad}}\right|}{c}. 
    \label{eq:teq}
\end{equation}
The equilibrium gas temperature, therefore, depends on $\left|\vec{F_\mathrm{rad}}\right|$ and $T_\mathrm{rad}$.

If the radiation source is a star with effective temperature $T_*$ and radius $R_*$, located at distance $r\gg R_*$ from the point under consideration, the flux drops with distance as 
\begin{equation}
    \left|\vec{F_\mathrm{rad}}\right| = \frac{ac}{4} T_*^4 \times \left(\frac{R_*}{r}\right)^2,
    \label{eq:flux}
\end{equation}
if extinction is neglected.
Then, the equilibrium gas temperature satisfies
\begin{equation}
    T_\mathrm{eq} \times \left( \frac{\kappa_\mathrm{P} \left( T_\mathrm{eq} \right) }{ \kappa_\mathrm{P} \left( T_\mathrm{eq}, T_* \right) } \right)^{1/4} = T_* \times \left( \frac{R_*}{2r} \right)^{1/2}.
    \label{eq:teqsimple}
\end{equation}

\begin{figure}[htb] 
  \begin{center} 
    \includegraphics[width=1.0\linewidth, angle=0]{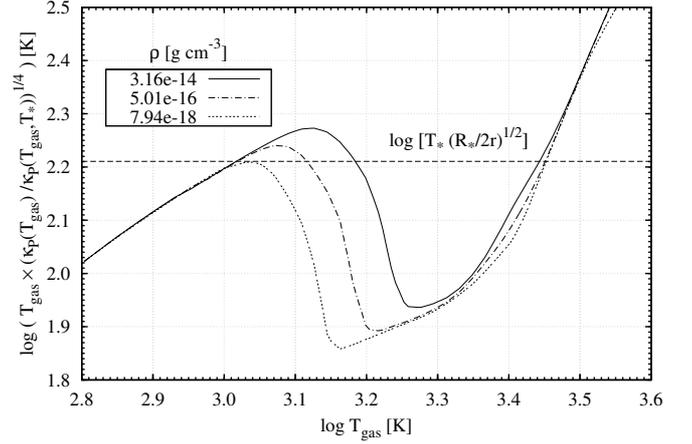}
    \caption{ Equilibrium temperature degeneracy in an optically thin gas. 
    The radiation source is a star with $T_* = 10^{4}\mbox{ K}$, $R_* = 1.7\mbox{ R}_{\odot}$ at $r = 15\mbox{ AU}$. 
    The temperatures at which the curves cross the dashed horizontal line are equilibrium temperatures. 
    See text for more details.
    }\label{fig:TDI}
    \end{center}
\end{figure}

For fixed $T_*$, $R_*$, and $r$, the right-hand side is a constant and the equilibrium temperature is defined by the ratio $\kappa_\mathrm{P}\left(T_\mathrm{gas}\right)/\kappa_\mathrm{P}\left(T_\mathrm{gas}, T_\mathrm{rad}\right)$. 
For a given density this ratio can vary with $T_\mathrm{gas}$ such that multiple solutions to Eq.~\eqref{eq:teqsimple} are possible. 
Figure~\ref{fig:TDI} illustrates this sort of degeneracy.
We consider a star with $T_* = 10^{4}\mbox{ K}$, $R_* = 1{.}7\mbox{ R}_{\odot}$, choose three different densities at $r = 15\mbox{ AU}$ ($3\times10^{-14}, 5\times10^{-16}, 8\times10^{-18}\mbox{ g\,cm}^{-3}$), and plot the left-hand side of Eq.~\eqref{eq:teqsimple} as a function of gas temperature. 
Temperatures at which the curves cross the dashed horizontal line, e.g. the right-hand side of Eq.~\eqref{eq:teqsimple}, are equilibrium temperatures. 
As is seen in the plot, there are multiple solutions for densities above $\approx8\times10^{-18}\mbox{ g\,cm}^{-3}$ and only a unique solution for lower densities. 
Getting closer to the star (shifting the dashed horizontal line upwards) would increase the value of this density threshold\footnote{At very low densities non-equilibrium chemistry may change the abundances of some species affecting the mean opacity as well. 
The degeneracy will only occur provided there is an inverse in the slope of the cooling (or heating) rate with respect to temperature.}.

In case of degeneracy, the first and the third solutions are stable, the second one is unstable.
Stability is given for positive gradients: $\partial_{T_{gas}}\left( T_\mathrm{gas} \times \left( \kappa_\mathrm{P}\left(T_\mathrm{gas}\right) / \kappa_\mathrm{P}\left(T_\mathrm{gas}, T_\mathrm{rad}\right) \right)^{1/4} \right) > 0$. 
Near a stable equilibrium point, gas with temperature $T_\mathrm{gas} < T_\mathrm{eq}$ will be effectively heated up to the equilibrium temperature and vice versa: gas with temperature $T_\mathrm{gas} > T_\mathrm{eq}$ will be effectively cooled down to the equilibrium temperature. 
The opposite happens around an unstable equilibrium point: runaway heating or cooling brings the temperature apart from that value to the corresponding stable solution. 
This process is known as radiative instability. 

\subsection{Equilibrium gas temperature as a function of radiation temperature}\label{sec:OTG:TGTR}
We can illustrate the degeneracy in another complementary way.
Namely, if we keep constant both the gas density and the geometric attenuation factor $(R_\mathrm{rad}/2r)^{1/2} = A$, we get $T_\mathrm{eq}$ in dependence of $T_\mathrm{rad}$:
\begin{equation}
    T_\mathrm{eq} = A T_\mathrm{rad}\times \left(\dfrac{\kappa_\mathrm{P}\left(T_\mathrm{gas}, T_\mathrm{rad}\right)}{\kappa_\mathrm{P}\left(T_\mathrm{gas}\right)}\right)^{1/4}.
    \label{eq:tgastrad}
\end{equation}
This equation has to be solved implicitly because $\kappa_\mathrm{P}\left(T_\mathrm{gas}, T_\mathrm{rad}\right)$ and $\kappa_\mathrm{P}$ are functions of $T_\mathrm{eq}$. 
\begin{figure}[!htb] 
  \begin{center} 
    \includegraphics[width=1.0\linewidth, angle=0]{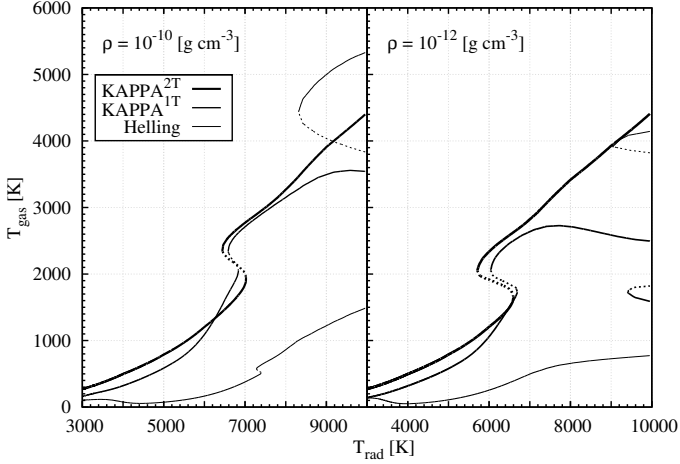}
    \caption{ Equilibrium temperature degeneracy in an optically thin gas. 
    The attenuation factor is set at $6{.}3\times10^{-2}$, which corresponds to $r \approx 127 R_*$.
    Each panel refers to a different density. 
    Left panel: $10^{-10}\mbox{ g\,cm}^{-3}$, right panel: $10^{-12}\mbox{ g\,cm}^{-3}$. 
    Each curve corresponds to a different opacity data source: the thickest line -- this work, using $\kappa_\mathrm{P}\left(T_\mathrm{gas}, T_\mathrm{rad}\right)$; the intermediate-thick line -- this work, using $\kappa(T_\mathrm{rad})$ instead of $\kappa_\mathrm{P}\left(T_\mathrm{gas}, T_\mathrm{rad}\right)$; the thinnest line -- \citet{HEL00}, using $\kappa(T_\mathrm{rad})$.
    The dashed parts of the lines mark unstable solutions (see text). 
    }\label{fig:TDII} 
    \end{center}
\end{figure}
We plot in Fig.~\ref{fig:TDII} equilibrium gas temperature versus radiation temperature for different opacity data. 
The attenuation factor $A = 6{.}3\times10^{-2}$ corresponds to $r \approx 127 R_*$ ($r = 1\mbox{ AU}$ for $R_* = 1{.}7\mbox{ R}_\odot$).  
Each panel corresponds to a different density. 
As can be seen, the degeneracy is not an exclusive feature of the new opacity data: it also appears when using opacity from \citet{HEL00}. 
As soon as the unstable solutions appear between the stable ones, limit cycles in the temperature profile are possible if one considers time-dependent radiative transport.


\section{Discussion}\label{sec:DISCUS}
Our calculations of mean gas opacities are based on three major simplifying assumptions: equilibrium chemistry, local thermodynamic equilibrium, and an ideal gas equation of state. 
The data presented in this paper is for zero value of the microturbulence parameter. 
Not all of these conditions may be fulfilled in a circumstellar environment.
Regions that are optically thin to stellar radiation could be affected by the change in abundances due to photochemistry. 
Whereas turbulent broadening can be easily taken into account via its parametrisation, giving up any of the three simplifications mentioned above could have entailed a noticeable increase in computational time. 
Equilibrium chemistry assumption should also be viewed as a first-order approximation from the perspective of utilising the advanced gas opacities in global collapse numerical modelling. 
Additional orders is a matter for further investigations. 
Moreover, current studies disagree in Planck mean gas opacities even though they employ a similar simplified approach. 
Accordingly, getting beyond the outlined scope would have partially eliminated one of the intentions of this paper: to demonstrate that insufficient frequency sampling is the true reason for this divergence. 
Thus, the framework outlined above is suitable for this initiating study.
The comparison to \citet{FERG05} and \citet{FREED08} verifies our data. 
Our Planck means disagree with \citet{HEL00} because of the insufficient frequency sampling employed in this (see Sect.~\ref{sec:RES:COMPAR}). 

The equilibrium gas-temperature degeneracy stems from radiative instability caused by e.g. molecule formation \citep{SCHIRR03}. 
The radiative cooling rate of an optically thin gas increases with its temperature if there is no phase transitions or radiative instability. 
If the gas temperature is around the point where at least one of its constituents experiences a transition from atomic to molecular phase, the gas can start to cool more efficiently because the excited molecules will radiate the thermal energy away (heating from the binding energy release competes with this process). 
After passing the transition, the derivative of the cooling rate with respect to gas temperature regains a positive value. 
This kind of a turnover in heating/cooling rate, being put into the framework of time evolution, leads to limit cycle instabilities that occur in cataclysmic variable events \citep[where it is due to hydrogen ionisation; see ][]{MEYER82}. 
Equilibrium temperature degeneracy facilitates spatial coexistence of cool and warm phases \citep[][]{SCHIRR03}, which is consequently reflected in the observable structure of the medium \citep[as is likely the case in TT Cyg; see ][]{OLOF00}. 

Equilibrium atomic-gas temperature determinations employing elaborate statistical equilibrium calculations were performed for simplified cases: \citet{WEY60} considered a compressible atomic gas under fixed heating rate in application to stellar chromospheres; \citet{WIL67} considered an incompressible atomic gas excited by ultraviolet synchrotron photons. 
Both studies revealed steep gradients in the equilibrium temperature profiles. 

A detailed discussion of all the heating and cooling mechanisms in a gas mixture is beyond the scope and intention of this paper. 
We therefore refer the reader to the excellent work by \cite{KAMP01}, where the authors considered dusty gas and additional gas heating from the photoelectrons extracted out of the dust particles. 

To couple a calculation of a detailed ionisation and chemical equilibrium to a radiative hydrodynamics numerical modelling within a framework of global collapse would demand a great deal of computational time even with modern computational facilities, which is why we consider radiative equilibrium in terms of mean opacity.
The frequency averaging is further justified by uncertainties and incompleteness of the present atomic and molecular data \citep[e.g.][]{KUR02}.

We highlight the impact this degeneracy has for the radiative transfer module described in \citet{KUI10}. 
The module includes an iterative temperature update scheme based on equation Eq.~(16) in \citet{KUI10}, 
\begin{equation}
    aT^4 = E_\mathrm{R} + \frac{\kappa_\mathrm{P}(T_\mathrm{rad})}{\kappa_\mathrm{P}(T)} \frac{\left|\vec{F_\mathrm{rad}}\right|}{c}, 
    \label{eq:kuiper}
\end{equation}
where $E_\mathrm{R}$ is the diffusive part of the radiation energy density, a term that dominates at high optical depths; 
the net radiation flux, $\left|\vec{F_\mathrm{rad}}\right|$, is evaluated at the point under consideration; 
and $c$ is the speed of light. 
In the case of gas, $\kappa_\mathrm{P}\left(T_\mathrm{gas}, T_\mathrm{rad}\right)$ should be used instead of $\kappa_\mathrm{P}(T_\mathrm{rad})$. 
In an optically thick medium, the local flux of radiation is dominated by its diffusive thermal part. 
The effective opaqueness to this radiation is expressed in terms of the Rosseland average (Eq.~\eqref{eq:ross}) and the radiative transfer equation for the net flux is treated in a flux-limited diffusion approach \citep{LP81}. 
Contrarily, in an optically thin region, the second irradiation term on the right-hand-side of Eq.~\eqref{eq:kuiper} dominates over the $E_\mathrm{R}$. 
Then, $\left|\vec{F_\mathrm{rad}}\right|$ can be substituted by the flux from the outer source (a star), $\left|\vec{F_\mathrm{rad}}\right| = L_*/(4\pi r^{2}) = (R_*/2r)^{2} \sigma T_*^{4}$, which turns Eq.~\eqref{eq:kuiper} into Eq.~\eqref{eq:teqsimple}.  
In the case of degeneracy, the iterative scheme is unsuited to finding a unique solution. 
To properly determine the gas temperature in optically thin parts, we propose taking into account its time-dependent evolution.


\section{Summary}\label{sec:SUM}
We calculated a new set of mean gas opacities assuming LTE and an ideal gas equation of state with a composition obtained from equilibrium chemistry for given atomic abundances.  
We will make all the tables freely available at the Strasbourg astronomical Data Center (the CDS, \href{http://cdsweb.u-strasbg.fr/}{http://cdsweb.u-strasbg.fr/}), as well as at \href{http://www.mpia-hd.mpg.de/homes/malygin}{http://www.mpia.de/\string~malygin}. 
For the purpose we used the publicly available code DFSYNTHE \citep{CAST05}. 
We covered the density and temperature ranges relevant for circumstellar environments. 
The means are tabulated in form of the Rosseland and the two-temperature Planck frequency averages on a grid of gas temperatures, gas densities (and radiation temperatures in the case of two-temperature Planck mean), and for three different metallicities: $\mathrm{[Me/H]}=-0{.}3,0{.}0,\mathrm{ and}+0{.}3$. 
We argue that the two-temperature Planck mean is essential for proper treatment of radiative energy balance in optically thin circumstellar gaseous environments. 

The part of the data that allows for comparison to other studies agrees with \citet{FERG05}.
We therefore conclude that our calculations are in line with the contemporary high-frequency resolution studies, which justifies that the convergence in gas opacity determinations has been achieved recently. 
The low-temperature - high-pressure part of our data provides a lower limit to the true opacity of a solar-composition dust-free gas, as inferred from the comparison to \citet{FREED08}.
We claim the new opacities may be profitably utilised for gas temperature determination 
in gaseous circumstellar environments. 
These would primarily correspond to the inner parts and atmospheres of accretion disks around forming stars as well as of passive disks around young stellar objects, circumstellar envelopes of pulsating stars, or upper irradiated parts of planetary atmospheres. 
In general, our data apply to any dust-depleted gaseous medium with a suitable range of thermodynamic parameters. 
 
Solving radiative equilibrium in an optically thin gas may yield more than one equilibrium temperature solution. 
This occurs because of radiative instability owing to transitions between molecular and atomic phases in gas. 
From the perspective of temporal evolution, the radiative instability may lead to limit cycles in the equilibrium temperature profile. 
Hence, the embedded gaseous cocoons around forming stars may have coexisting different-temperature phases similar to the ISM large-scale structure. 
In addition, abrupt temporal variations in temperature of the falling gas are expected to occur during star formation.
This can affect the accretion rate, similarly to how it happens in cataclysmic variables \citep{MEYER82}. 
The equilibrium temperature structure and its evolution will clearly affect the appearance of the dust-free interiors, although the question of how it alters the secular evolution of the host system (a massive proto-star) is yet to be addressed.  

The degeneracy of the equilibrium temperature has at least one important numerical corollary: the radiative transfer modules relying on iterative temperature update schemes will not, in general, be able to provide convergence of the solution.
The methods tracking the evolution of the temperature with time are therefore requisite for such modules.

We obtained the equilibrium temperature degeneracy from consideration within a simplified framework: equilibrium chemistry determines the relative abundances of the compounds, local thermodynamic equilibrium defines the level populations, and the medium is optically thin. 
We have not yet investigated how the inclusion of non-equilibrium chemistry, non-LTE effects, frequency-resolved radiative transport, and/or other means of energy transport (e.g. convection) would affect the degeneracy. 
All of these are starting points for future investigations.

\begin{acknowledgements} 
    This research has been supported by the International Max-Planck Research School for Astronomy and Cosmic Physics at the University of Heidelberg (IMPRS-HD).
    MM thanks F.~Castelli and R.~L.~Kurucz for creating and sharing the opacity code with all the necessary descriptions; R.~Freedman for a kind assignation of the data obtained by his group; Michael Maseda for comprehensive reading of the script, which helped to improve it a lot; Yakiv Pavlenko, Tristan Hayfield, Paul Molli\`ere, Gabriel-Dominique Marleau and Mario Flock for fruitful discussions on various aspects of this research. 
\end{acknowledgements}

\bibliographystyle{aa}
\bibliography{aa_2014_23768-Malygin_et_al-MGOfCEaETD}

\end{document}